\documentclass[pre,twocolumn,floatfix,superscriptaddress,showpacs,preprintnumbers]{revtex4}

\usepackage{graphicx}
\usepackage{epsfig}
\usepackage{amsmath}

\begin{document}

\preprint{}

\title{Current Reversals and Synchronization in Coupled Ratchets}

\author{U.~E.~Vincent}
\email[Corresponding author:]{ u.vincent@tu-clausthal.de}
\affiliation{Department of Physics, Lancaster University, Lancaster LA1 4YB, United Kingdom}
\affiliation{Institut f\"ur Theoretische Physik, Technische Universit\"at Clausthal, Arnold-Sommerfeld Str. 6, 38678 Clausthal-Zellerfeld, Germany}
\author{A.~Kenfack}
\affiliation{Physikalische und Theoretische Chemie, Institut f\"ur Chemie und Biochemie,
Freie Universit\"at, Takustr. 3, 14195 Berlin, Germany.}
\author{D.~V. ~Senthilkumar}
\affiliation{Center for Dynamics of Complex Systems, University of Potsdam, 14469 Potsdam, Germany}
\affiliation{Postdam Institute for Climate Impact Research, 14473 Potsdam, Germany}
\author{D. ~Mayer}
\affiliation{Institut f\"ur Theoretische Physik, Technische Universit\"at Clausthal, Arnold-Sommerfeld Str. 6, 38678 Clausthal-Zellerfeld, Germany}
\author{J. ~Kurths}
\affiliation{Postdam Institute for Climate Impact Research, 14473 Potsdam, Germany}
\affiliation{Institut f\"ur Physik, Humboldt Universit\"at zu Berlin, Newtonstr. 15, 12489 Berlin, Germany}
\date{\today}

\begin{abstract}
Current reversal is an intriguing phenomenon that has been central to recent experimental and theoretical investigations of transport based on ratchet mechanism. By considering a system of two interacting ratchets, we demonstrate how the coupling can be used to control the reversals. In particular, we find that current reversal that exists in a single driven ratchet system can ultimately be eliminated with the presence of a second ratchet. For specific coupling strengths a current-reversal free regime has been detected. Furthermore, in the fully synchronized state characterized by the coupling threshold $k_{th}$, a specific driving amplitude $a_{opt}$ is found for which the transport is optimum.

\end{abstract}
\maketitle
Transport phenomena and, particularly, directed transport occur in many situations ranging from physical systems to chemical and biological systems. Some recent research interest in transport problems is related to ratchet physics where unbiased, noise-induced transport occurs away from thermal equilibrium as a result of the action of Brownian motors~\cite{Reimann2002,Hanggi2005,Astumian1998}. Brownian motors, especially ``ratchet'' models, have been widely investigated partly due to the challenge to describe and control mechanisms of fundamental biological processes at both the cell level (e.g. transport in ion channels) and body level (muscle operations)~\cite{Hanggi1996}. Another motivation is derived from recent advances in technology wherein devices for guiding tiny particles on nano/micro scales are sought; these include particle separation techniques, smoothing of atomic surfaces during electromigration and control of the motion of vortices in superconductors~\cite{Hanggi2005,Linke2002}. Remarkably, experimental realizations of some of these practical systems have been reported~\cite{Villegas2003,Matthias2003,Siwy2002}.

In this framework, two basic types of ratchet models have commonly been employed, namely: (i) the rocking ratchet, in which the particle is subject to an unbiased external force with or without additive noise, and (ii) the flashing ratchet, in which the particle is periodically kicked. The vast majority of these models are overdamped where the noise plays a vital role in the transport process. However, recent studies have shown that the role of noise can be replaced by deterministic chaos induced by the inertial term~\cite{Jung1996}. In such inertial ratchets, the issue of current reversal has been carefully investigated~\cite{Mateos2000,Barbi2000,Mateos2002,Mateos2003,Kenfack2007}. Moreover, Hamiltonian ratchets have recently seen a breakthrough in the ratchet community~\cite{Schanz2001}. Here the noise and particularly the dissipation are absent, thereby allowing these systems to preserve their full coherence. Hamitonian ratchets owe their merit to the first experimental realization of the quantum ratchet potential~\cite{Ritt2006, Salger2007} which has been a very good motivation for more theoretical as well as experimental works. In this context, higher order quantum resonances (a regime of very fast and directed transport) have been found with atoms~\cite{Kenfack2008}, and directed transport of atoms has quite recently been experimentally achieved\cite{Salger2009}.

In the description of the model transport in general, attention has mostly been paid to single particle ratchets. However many such systems coexist in numbers and do work in cooperation. For instance, molecular motors do not operate as a single particle but in groups - the most prominent example being the actin-myosin system in muscles~\cite{Alberts2002}. For this reason, the relevance of many interacting particles and possible effects of collective behavior, e.g. transport enhancement, current inversion, clustering, synchronization and spontaneous current among others have mostly been considered in the overdamped case (\cite{Cilla2001} and references therein). Besides, very little has been done on coupled underdamped ratchets~\cite{Kostur2005,Chen2005,Vincent2007} which is here the model of interest. This is due to the additional complexity induced by inertial terms and also to the presence of the dissipation. In a very recent paper on this issue of coupled ratchets, it has been shown that coupling may alter significantly the intensity and direction of the net rectified motion~\cite{Savelev2003}.

Thus, in this paper, we study dynamics of two coupled driven underdamped ratchets and show how the coupling can be used to control current reversals. The full synchronized state observerd and characteriszed by the coupling threshod $k_{th}$, is preceded by several sudden changes in the current and particularly current reversals. It is precisely in that state that we found a specific value of the driving amplitude $a_{opt}$ for which the transport is optimum. Furthermore, we report that by tuning the coupling parameter, a regime of current reversal free can be obtained. We note in passing that when the interaction between coupled superconducting Josephson devices is sufficiently strong to induce synchrony, the resulting synchronized dynamics could give rise to large power output~\cite{Blais2003,Nishanen2007}.

The model we are interested in is made of two rocking ratchets \cite{Mateos2000}, symmetrically perturbed by means of an elastic coupling. Its dynamics (in dimensionless form) can be described by
\begin{equation}
        \ddot x_i + b \dot x_i + \frac{dV(x_1,x_2)}{dx_i} = a \cos(\omega t) ~~~(i=1,2),
\label{eq(1)}
\end{equation}
where the normalized time $t$ is taken in  units of the small resonant frequency ${\omega_0}^{-1}$ of the system; $a$ and $\omega $ represent the amplitude and the frequency of the driving, respectively; and $b$ the damping parameter. Here  $V(x_1,x_2)$ is the perturbed two-dimensional ratchet potential given as:
\begin{equation}
V(x_1,x_2) = 2C-\frac{1}{4\pi^2 \delta}\left[ \phi(x_1)+\phi(x_2)\right]+ \frac{k}{2}{(x_1-x_2)}^{2},
\label{pot2}
\end{equation}
where $\phi(x_{1,2})=\sin 2\pi (x_{1,2}-x_0)+0.25\sin 4\pi(x_{1,2}-x_0),$ and $k$ is the coupling strength which determines the dynamics and hence the transport properties of Eq. (\ref{eq(1)}). The constant $C$ is merely introduced for a reference frame purposes. And to enforce this potential to take zero at the origin, that is $V(0,0)=0$, this constant must be given by $2C=-(\sin 2\pi x_0+0.25\sin 4\pi x_0)/4\pi^2\delta$. In this case for  $x_0=0.82$ and $\delta=1.614324$, $C\approx 0.0173$. Also, $b=0.1$ and $\omega=0.67$ are kept constant throughout. The system Eq.~\ref{eq(1)} models to some extent the well known Frenkel-kontorova system~\cite{braun2004}, with only two elastically coupled particles, which has recently been extensively used to study directed
transport~\cite{csahok1997,porto2000,zheng2001,flach2002}. Along these lines experiments have been successfully carried out with a circuit of parallel Josephson junctions array~\cite{trias2000}.

\begin{figure}
\includegraphics[width=8cm,height=8cm]{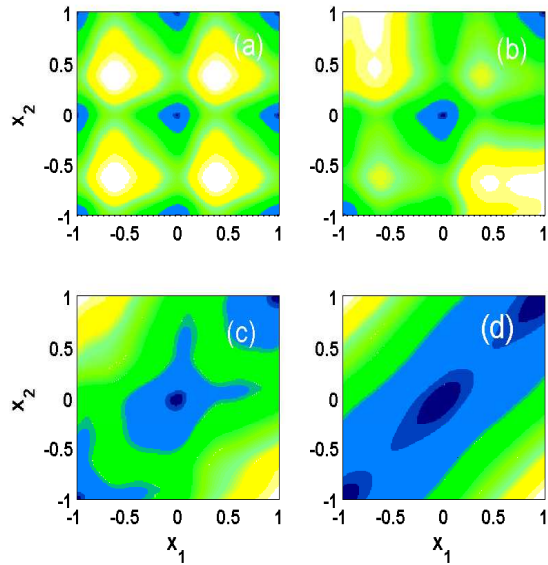}
\caption{(Color online) Equipotential contours plot of $V(x_1,x_2)$ with colors growing from dark-blue (minima), through yellow/gray (intermediate) to white (maxima): (a) No interaction, $k = 0$, (b) weak coupling, $k = 0.05$, (c) moderate coupling, $k = 0.15$ and (d) strong coupling, $k = 1.0$.}
\label{Fig1}
\end{figure}

The 2D ratchet potential (\ref{pot2}) is shown in Fig.~\ref{Fig1} for four different values of the coupling strengths. The minima and maxima of the potential are marked with dark-blue and white colors, respectively. It is to be noted that as $k$ is increased, the heights of the potential $V(x_1,x_2)$ move outward, opening up a valley along the diagonal, in which the two interacting ratchets may most likely share.

\begin{figure}
\includegraphics[width=8.0cm,height=9cm]{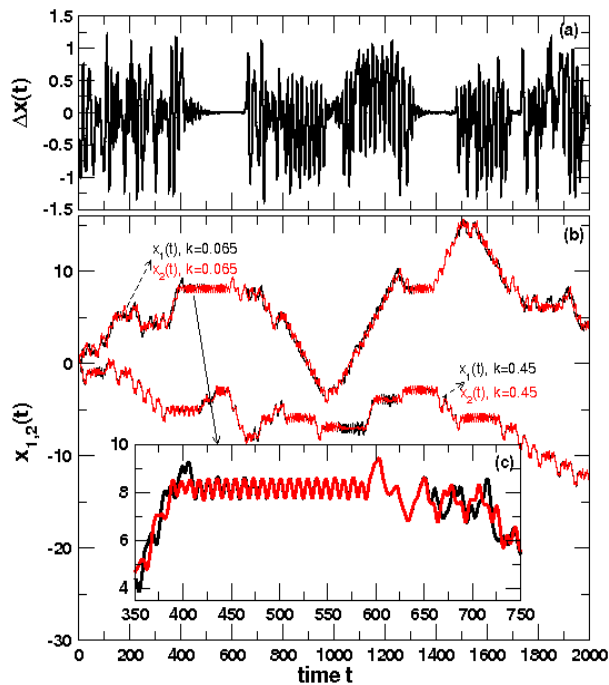}
\caption{(Color online) Typical on-off intermittency for $a=0.0809472$
(a) of the error state $\Delta x(t) = x_1(t)-x_2(t)$ for $k = 0.065$,
(b) of trajectories $x_1(t)$ (black) and $x_2(t)$ (red/gray) for $k=0.065$ and $k = 0.45$  as indicated with dashed arrows, and (c) of the enlarged
portion of (b) for $k = 0.065$.}
\label{Fig2}
\end{figure}

\begin{figure}
\includegraphics[width=8.0cm,height=9cm]{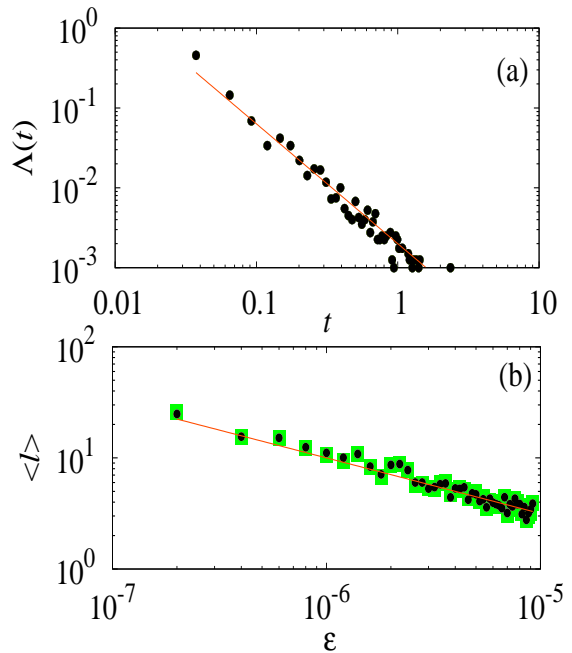}
\caption{(Color online) (a) Distribution of laminar phases $\Lambda(t)$ of $\Delta x(t)$  (black circles), satisfying a $-3/2$ power law scaling typical of on-off intermittency and (b) average laminar lengths with varying parameter $a$ satisfying the scaling law $\langle l \rangle ~ \propto \epsilon^{-0.5}$ with $\epsilon = a-a_c$ and $a_c=0.0809474$ [$x_1$ (black circles) and $x_2$ (green/gray squares)] showing type-I intermittency. Here $k = 0.065$ and the lines (red/gray) are corresponding fits.}
\label{Fig3}
\end{figure}

Fig.~\ref{Fig2} displays the behavior of a single trajectory analysis. Here, on-off intermittency~\cite{Fujisaka1985} can clearly be observed by the error state $\Delta x(t)=x_2(t)-x_1(t)$ sketched in Fig.~\ref{Fig2}(a). Fig.~\ref{Fig2}(b) shows typical trajectories of the system along which asynchronous motions erratically alternate with quasiperiodic ones. This scenario is illustrated in the enlargement portion of Fig.~\ref{Fig2}(b), say Fig.~\ref{Fig2}(c). To confirm and quantitatively characterize the intermittency, we have plotted in Fig.~\ref{Fig3}, for $k=0.065$, the probability distribution of the laminar phases $\Lambda(t)$ of $\Delta x(t)$ and the average laminar lengths $\langle l \rangle$ of the two trajectories $x_{1}(t)$ and $x_{2}(t)$ as function of $\epsilon = a - a_{c}$. Here $a_c = 0.0809474$ is a critical driven amplitude at which each subsystem $x_1$ or $x_2$ for $k=0$ undergoes a bifurcation from chaotic to periodic regimes, which is associated to current reversal~\cite{Mateos2002}. In Fig. \ref{Fig3}(a), the collective dynamics for the coupled ratchets shows a $-3/2$ power law scaling, typical of on-off intermittency, while each subsystem dynamics, $x_1$ or $x_2$, exhibits type-I intermittency similar to the single ratchet dynamics~\cite{Son2003} with a $-1/2$ power law scaling as shown in Fig. \ref{Fig3}(b). At this level, the observed intermittency is a clear indication of the complexity of the dynamics and suggests to account for statistical calculations for any observable of the system. The above picture based on a single trajectory analysis is exact for single attractor systems and may turn out to be misleading for irregular ones, where periodic and chaotic attractors could co-exist.

Next, we explore the dynamics of the coupled ratchets (\ref{eq(1)}) as $k$ is varied and considering that single trajectory dynamics will not suffice for a highly chaotic system, all observables have to be averaged out over a large number of trajectories generated from the entire space $[-1,1]\times[-1,1]$ which is the unit cell of the resulting periodic structure. Here we make use of two important indicators, namely the error state $\eta$ as a good measure of the synchronization and the current $J$ as the transport quantifier. For a long time dynamics $T$, the error state for a given trajectory is given by:
\begin{equation}
\eta_j= \frac{1}{T}{\int_0}^T\left[({x_2}^{(j)}-{x_1}^{(j)})^2+{({\dot x_2}^{(j)}-{\dot x_1}^{(j)})}^2\right]^{1/2} dt,
\label{eqn3}
\end{equation}
with the full error $\eta=N^{-1}\sum_{j=1}^{N}\eta_j$, evaluated over the total number $N$ of trajectories. On the other hand, the current in a subsystem $(i=1,2)$ is defined as follows:
\begin{equation}
J_i=\frac{1}{M-{n_c}}\frac{1}{N}\sum^{M}_{l=n_{c}}\sum^{N}_{j=1}{\dot x_i}^{(j)}(t_{l}) ~~~~ (i=1,2).
\label{eqn4}
\end{equation}
where $N$ is the total number of trajectories, $t_l$ a given observation time and $M$ the total number of observations. This gives the average velocity, which is then further time-averaged over the number of  observations $M$ - $n_c$.  Here $n_c$ is an empirically obtained cut-off accounting for the transient effect, such that a converged current is obtained~\cite{Kenfack2007}.

\begin{figure}
\includegraphics[width=8.0cm,height=11.0cm]{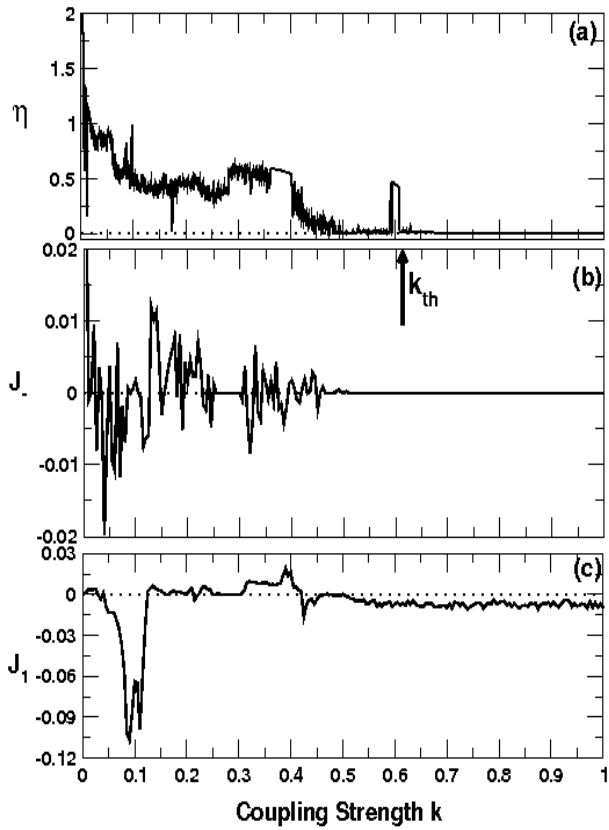}
\caption{ Transition to full synchronization, as function of the coupling strength $k$ and for $a=0.0809472< a_c$, indicated by (a) the average error dynamics $\eta$ and (b) the current $J_- = J_1-J_2$. In the same coupling range, (c) the current $J_1$ is plotted showing regions of zero current and of current reversals. Here $k_{th}$ is the coupling threshold for full synchronization.}
\label{Fig4}
\end{figure}

With the driving amplitude $a=0.0809472< a_c$, each independent system $(k=0)$ exhibits chaotic dynamics~\cite{Mateos2000}. Fig.~\ref{Fig4}(a) displays $\eta$ as a function of $k$. Above the threshold $k > k_{th} \approx 0.576$, $\eta$ approaches zero, indicating a fully synchronized state. In Figs.~\ref{Fig4}(b) and ~\ref{Fig4}(c) we observe the global dependence of the currents $J_- = J_1 - J_2$ and $J_1$, respectively, on the degree of synchronization. Prior to the synchronized state, $J_-$ fluctuates around zero and when a full synchrony is achieved, $J_-$ is identically zero. Notice that the nonzero $J_-$ occur at weaker coupling $k<<k_{th}$. In this case, the two particles interact less and may predominantly evolve in the same direction with slightly different velocities. However in the narrow band $0.259 \le k \le 0.295$, $J_-$ is identically zero; this zero current is not associated with full synchrony, but rather to no directed transport by the individual systems (see Fig.~\ref{Fig4}(c)). Such a situation, already reported in a single ratchet system~\cite{Mateos2002}, occurs when the average velocity of the particle is zero. This implies that in this case, in the present system, each particle does not necessarily reside completely in a quasiperiodic state, but may experience chaotic bursts for which the average velocity is typically zero.

\begin{figure}
\includegraphics[width=7.50cm,height=7.0cm]{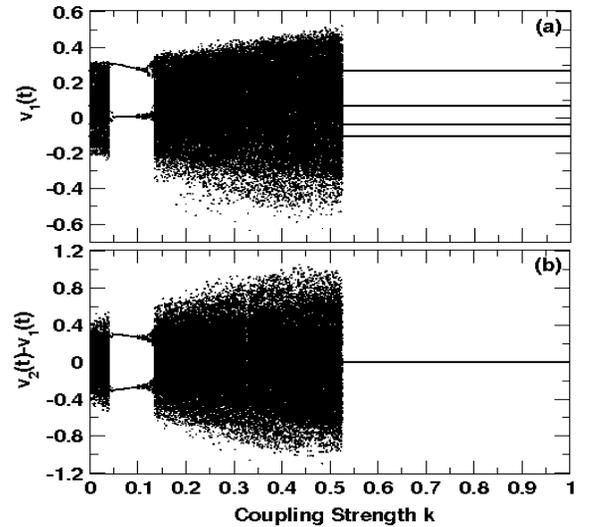}
\caption{The bifurcation diagrams for $a=0.0809472$ as function of the coupling strength $k$. The transition from the intermittent chaotic regime to a period $4$ orbit (a) $v_1(t)={\dot x_1}(t)$, happens at the synchronization threshold $k\ge k_{th}~\approx 0.576$ (b) $v_2(t)-v_1(t)$.}
\label{Fig5}
\end{figure}

\begin{figure}
\includegraphics[width=7.50cm,height=4.0cm]{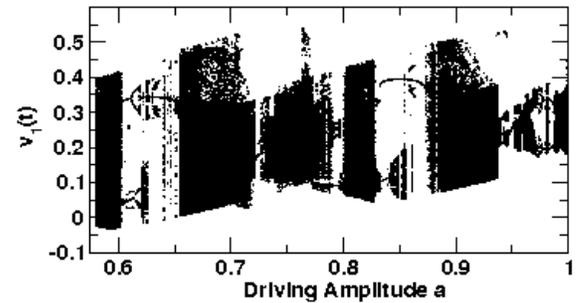}
\caption{Bifurcation diagram for large amplitudes of the driving force $a$, for $k=0$, showing several windows of chaos as well as quasiperiodic ones.} \label{Fig6}
\end{figure}

The spikes in $J_1$ with or without current-reversals reveal some sudden changes in the corresponding bifurcation diagrams. In Fig.~\ref{Fig5} we show two bifurcation diagrams, as function of $k$, corresponding to the velocities (a)  $v_1(t)$ and (b) $v_2(t)-v_1(t)$ in the same coupling range as in Fig.~\ref{Fig4}, where $v_i(t)={\dot x_i}(t)$, $i=1,2$. The transition to full synchrony described in Fig.~\ref{Fig4} is clearly reflected in the underlying dynamical behavior. First, a sudden change occurs at $k \approx 0.04$ during which a bifurcation from chaotic state to a period two (${\it P_2}$) window is detected. This bifurcation corresponds to the current-reversal (see Fig.~\ref{Fig4}(c)). Notice that the ${\it P_2}$ orbit remains stable in some range of the coupling strength, namely $0.04 \le k \le 0.13$; then undergoes a Hopf bifurcation when the strength of the interaction further increases; and a chaotic regime again shows up for a wide range of $k$. Next a sudden bifurcation takes place at a critical value $k_{th}$ at which the dynamics of the two ratchets become locked in complete synchronization as shown in Fig.~\ref{Fig5}(b) and current-reversal takes place, see Fig.~\ref{Fig4}(c). During this transition to the full synchrony ($k \ge k_{th}$), a period $4$ orbit is born (Fig.~\ref{Fig5}a). Note that the dynamics of a single ratchet ($k=0$) at stronger driving amplitudes $a$, is also very complex as can be seen from Fig.~\ref{Fig6}, with several windows of chaos separated with quasiperiodic ones. One may thus anticipate that the physics at weaker or stronger driving forces is qualitatively similar.


\begin{figure}
\includegraphics[width=7.50cm,height=7.0cm]{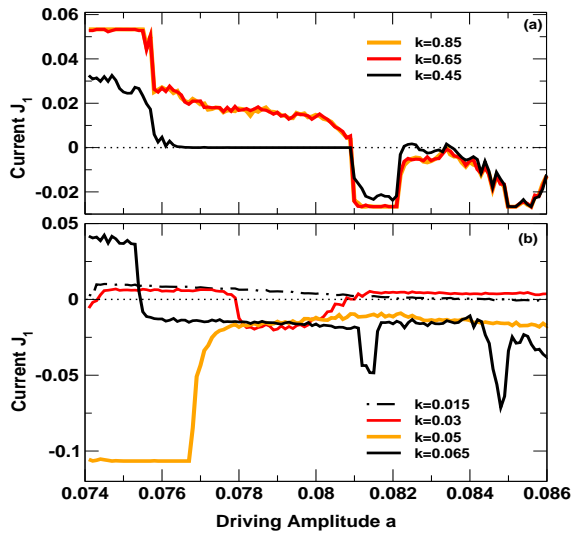}
\caption{(Color online) Current $J_1$ as function of the driving amplitude $a$ for (a) strong couplings $0.4 < k < 1$
and (b) weak couplings $0 < k < 0.07$ as indicated in each panel.}
\label{Fig7}
\end{figure}

The above observations allows us to pay special attention to controlling current reversals as these may happen to be undesirable as far as transport is concerned. The case $k=0$ corresponds to no interactions for which current-reversals have been observed in a single ratchet model~\cite{Mateos2000,Kenfack2007}. The strong coupling regime, where full synchronization is reached, corresponds identically to the current reversal observed in one single ratchet model, see Fig.~\ref{Fig7}(a). However as $k$ takes on smaller values, dramatic changes occur on current leading to the {\it rectification of the particles motions} - for instance, current reversal observed in a large window of $a\in (0.0808,0.0823)$, shown in Fig.~\ref{Fig7}(a), for strong couplings $k\in (0.4,1)$ is completely eliminated in Fig.~\ref{Fig7}(b) for weaker couplings $k\in (0,0.07)$. Likewise current reversals found for smaller values of $k$ are destroyed as $k$ increases, see Fig.~\ref{Fig4}(c). The system becomes totally reversal-free for example at $k=0.015$ and $k=0.05$ , see Fig.~\ref{Fig7}(b). This result clearly demonstrates the importance of the coupling strength over the full control of transport.

\begin{figure}
\includegraphics[width=7.50cm,height=4.0cm]{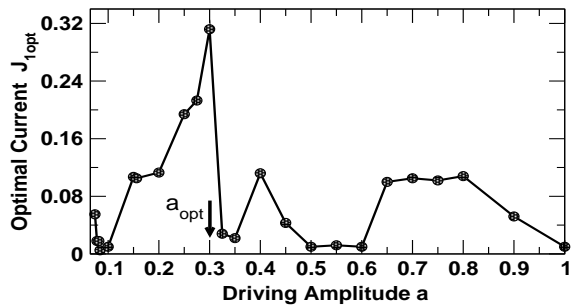}
\caption{Optimal current ${J_1}{opt}$ as function of the driving amplitude  $a$ over the coupling interval $0 \le k \le 1.0$. One clearly sees the \emph{transport suppression} for weak amplitudes ($a < 0.1$), and also for $0.5<a<0.6$, and \emph{transport enhancement} for other values of $a$. The peak optimal current is at a moderate driving force $a=a_{opt}\approx 0.3$.} \label{Fig8}
\end{figure}

At this point the question that may naturally arises is wether or not there are parameters for which the transport can be enhanced. For the entire driving amplitude range $a\in(0,1)$, we have systematically computed the current ${J_1}$ for the entire coupling strength range $k\in(0,1)$. The recorded optimal current ${J_1}_{opt}$, absolute value of ${J_1}$, is achieved coincidentally as the synchronization regime is reached, $k\ge k_{th}$, and remains constant throughout. We plot in Fig.~\ref{Fig8} for $k\in (0,1)$, the quantity ${J_1}_{opt}$ as function of the driving amplitude $a$. Here, we clearly identify regimes of \emph{transport enhancement} and \emph{suppression} triggered by synchronization.  In the weak forcing regime, typically $a < 0.1$, and also for $a\in(0.5,0.6)$, the current is suppressed; while the remaining forcing regime exhibits {\it optimum transport} at $a=a_{opt}\approx 0.3$. This picture, which is likely a good guide of the efficient transport, may be very interesting for experimental purposes.

To sum up, we have clearly shown that the dynamics of a single particle ratchet can be significantly modified when coupled elastically to a second one. We have thus made use of the coupling strength to systematically rectify the particle motion. In particular, current reversals observed in a single ratchet can completely be annealed by appropriately choosing the coupling strength. A regime of current-reversal free has thus been detected for specific lower coupling strengths. Then we found a coupling threshold $k_{th}$ for which the system is fully synchronized. Exploring parameters space, in this synchronized state, we demonstrated that the transport can either be enhanced or suppressed, depending on the driving amplitude $a$. In doing so, we were able to to find a specific driving amplitude $a_{opt}$ for which the optimum transport is achieved. These results clearly demonstrate the importance of the coupling strength and that of the driver over the full control of non equilibrium transports.

\section*{Acknowledgment:} UEV and DVS are supported by the Alexander von Humboldt Foundation, Germany. JK acknowledges the support from  EU  under project No. 240763 PHOCUS (FP7-ICT-2009-C). Comments and suggestions by the reviewers are immensely acknowledged.

\end{document}